\documentstyle[prl,tighten,aps,epsf,amstex,multicol,times,bbm]{revtex}
\begin{document}

\draft

\newcommand{\proofend}{\hfill\fbox\\ }

\title{Conditions for the local manipulation of
Gaussian states}

\author{Jens Eisert and Martin B. Plenio}

\address{QOLS, Blackett Laboratory, Imperial College of Science, Technology
and Medicine, London, SW7 2BW, UK}

\date{\today}
\maketitle

\begin{abstract}

We present a general necessary and sufficient criterion for the
possibility of a state transformation from one mixed Gaussian
state to another of a bi-partite continuous-variable system
with two modes. 
The class of operations that will be considered is 
the set of local Gaussian
completely positive trace-preserving maps.
\end{abstract}

\pacs{PACS-numbers: 03.67.-a, 03.65.Bz, 42.50.Lc}

\begin{multicols}{2}
\narrowtext Imagine a physical device that is able to manipulate
locally the state of a composite quantum system by actions on its
parts. Which state transformations could this device implement in
principle, abstracting from experimental imperfections? This
question is particularly important in the field of quantum
information theory \cite{Intro1}, which concerns itself with the
problem of whether a certain resource, e.g. an entangled quantum
system in a known state, could be used to accomplish an envisioned
task. To be more specific, one asks for mathematical conditions
that have to be met in order for a state transformation under
natural constraints to be possible.

Such a natural constraint is that the device can only implement
local quantum operations supplemented by classical communication
(LOCC), as many applications in quantum information science involve
spatially separated parties sharing entangled states. So far, when
investigating transformation criteria under LOCC all efforts have
been devoted to the case where the involved quantum systems
possess finite-dimensional Hilbert spaces as, e.g., qubit systems.
The widely acknowledged result of Ref.\ \cite{Nielsen} relates the
problem of the deterministic transformation between pure states by
means of LOCC to the mathematical theory of majorization. Based on
this insight a series of further results have been found
\cite{Jonathan,MixedRemark}.
%
While the constraint to general LOCC is natural
for low-dimensional systems, the 
situation is quite different for systems with
an infinite-dimensional Hilbert space, such as the modes of an
electromagnetic field.
The experimental operations that are typically available
are those that involve beam-splitters, phase shifters, and
squeezers together with the ability to prepare ancilla systems in
a standard state such as the vacuum. The class of states that can
be generated by these operations, and which is therefore
particularly relevant from an experimental point of view, is the
set of Gaussian states
\cite{Simon,Giedke,Symplectic,Lindblad}. Several 
properties of
entangled Gaussian states are already 
known. In particular, the problems of
distillability and separability of Gaussian states have been
investigated in great detail, and can actually be considered
solved \cite{Simon,Giedke}. However, 
the general 
question of the local
interconvertability between entangled
Gaussian states has not been addressed before.

This letter 
presents a first step towards finding
tools for deciding whether a desired transformation of Gaussian
states can be accomplished without the need of going through all
physical protocols, which can be an extremely tedious task. We
will present a general necessary and sufficient criterion for the
possibility of state transformations of a two-mode
continuous-variable system. The class of allowed operations is the
set of local Gaussian completely positive operations \cite{LOGd}, 
that is, those local operations that can be realized by means
of local joint symplectic transformations
on both the system and arbitrary appended ancilla 
systems that have been prepared in Gaussian states. This
set will be abbreviated as LOG, and 
the statement that a
transformation from a state $\rho$, pure or mixed,
to a state $\rho'$ is possible will be written as
\begin{equation}\nonumber
    \rho\longrightarrow \rho'\,\,\text{ under LOG}.
\end{equation}
In quantum optical systems this 
class of operations is the one that can
be realized (with present technology)
as a combination of
applications of 
beam splitters, phase shifts, and squeezers
together with the
possibility to append additional field modes locally. 

\begin{figure}
\centerline{
        \epsfxsize=6.5cm
       \epsfbox{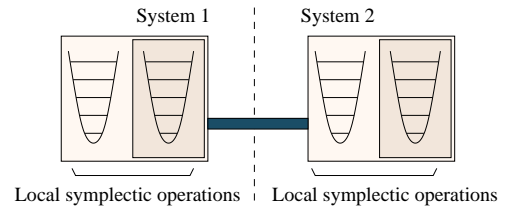}
}

\vspace{.2cm}
\caption{
Any local Gaussian completely positive map can be conceived as
a composition of a local 
joint symplectic transformation on both 
the system and additional
oscillators which have been
prepared in a Gaussian state 
and a partial trace operation with respect to the
additional oscillators.
}
\end{figure}

The physical system under consideration is a bi-partite
quantum system
with one canonical degree of freedom
each, such as
two modes of an electromagnetic field.
As in Refs.\ \cite{Giedke} such a system will be called
a
$1\times 1$-system, consisting of parts 1 and 2.
In order to exploit the elegant formalism
that is available
to describe Gaussian quantum states
\cite{Simon,Symplectic}
it is convenient to
group the Hermitian operators corresponding to
position and momentum in a vector,
${\mathbf{O}}=(X_1,P_1,X_2,P_2)$.
The canonical commutation relations (CCR)
can then be subsumed into
the skew symmetric block diagonal
$4\times 4$-matrix $\Sigma$ according to
     $[O_n, O_m]= i \Sigma_{n m}$,
$n,m=1,...,4$.
For a given state $\rho$, for which the second moments
exist, let the real $4 \times 4$-matrix
$\Gamma$ be defined as
\begin{eqnarray}\nonumber
    \Gamma_{n m}
    &=&
    2 \text{tr}\left[
    \rho \left(O_n-\langle O_n\rangle_{\rho} \right)
    \left(O_m-\langle O_m\rangle_{\rho}  \right)
    \right] -i \Sigma_{n m}
\end{eqnarray}
where $\langle O_n\rangle_{\rho}=\text{tr}[\rho O_n]$.
The matrix $\Gamma$
will be referred to as covariance matrix.
Not all symmetric $4\times 4 $-matrices
are legitimate covariance
matrices: the restriction that $\rho$ is a
state manifests itself
as the condition
$\Gamma - i \Sigma \geq 0$
for the covariance matrix, which is in fact a formulation 
of the uncertainty relations.
For Gaussian states \cite{Ga}
the covariance matrix together with the
mean values of the position and momentum operators are
sufficient to fully specify the
state.
The first moments, however,
are of no relevance for the issue of this paper,
because
they can always be made to vanish
by an
appropriate local translation in phase space.

We will now turn to the possible state transformations.
Of particular interest are
the linear transformations from
one set of canonical coordinates to another set
which leave the CCR invariant.
In a system with two canonical degrees of freedom
they
form the group of real 
symplectic transformations $Sp(4,\mathbbm{R})$
\cite{Symplectic}.
The group  $Sp(4 ,\mathbbm{R})$ consists of the real
$4\times 4$-matrices
$S$ obeying $S^T \Sigma S=\Sigma$;
the group $Sp(2N,\mathbbm{R})$ can be defined
in an analogous manner for $N$ canonical degrees
of freedom.
Under a symplectic transformation a
covariance matrix is transformed according to
$\Gamma\longmapsto  S^T \Gamma S$. On
the level of states it is accompanied by a
unitary operation $\rho\longmapsto U(S)\rho U(S)^{\dagger}$,
then called symplectic operation.
A local symplectic transformation is a matrix $S$
of the form $S=S_{1}\oplus S_{2}$, where $S_{1}, S_{2}\in Sp(2, \mathbbm{R})$.
The most general
LOG can now be conveived as a composition of
a joint symplectic transformation $S=S_{1}\oplus S_{2}$
with $S_{1},S_{2}\in Sp(2N+2, \mathbbm{R})$
on the original systems 1 and 2
and on two additional systems
with $N$ canonical degrees of
freedom each of which has been locally
prepared in a Gaussian state, and
a partial trace operation with respect to the
additional systems (see Fig.\ 1).

Any covariance matrix $\Gamma$ of a bi-partite $1\times 1$-system
can be written in block form as
\begin{eqnarray}\label{block}
    \Gamma= \left(
    \begin{array}{cc}
    A_1 & B\\
    B^T & A_2
    \end{array}
    \right),
\end{eqnarray}
where $A_1$,$A_2$, and $B$ are real $2\times 2$-matrices
\cite{Inv}.
One can uniquely
characterize the orbit $O(\Gamma)$
of $\Gamma$
with respect to local symplectic transformations
by a vector
$(\xi_1, \xi_2 ,\xi_3,\xi_4)\in{\mathbbm{R}}^{4}$,
the entries of which are given by
$\xi_1:= |A_1|^{1/2}\geq 1 $,
$\xi_2:= |A_2|^{1/2}\geq 1$, where $|.|$ denotes
the determinant.
$\xi_3$
and $\xi_4$ are the solutions of $\xi_3 \xi_4=|B|$,
    $\xi_3^2 + \xi_4^2 = (|B|^2 - |\Gamma| +
    |A_1|
    |A_2|)/(|A_1||A_2|)^{1/2}$,
such that
$\xi_{3}\geq | \xi_{4} |$.
It has been shown in Ref.\ \cite{Simon}
that
$\Gamma$ can always be
transformed into a covariance matrix
$S^{T} \Gamma S$ which is of `normal form'
by using an
appropriate local symplectic transformation $S$:
this means that $S^{T} \Gamma S$
is of the form of Eq.\ (\ref{block}),
but with
    $B= \text{diag}(\xi_{3} ,\xi_{4})$ and
    $A_{i}= \text{diag}(
    \xi_{i},\xi_{i})$, $i=1,2$.
%

Whether a transformation of a state $\rho$ to
a state $\rho'$ with respective covariance matrices $\Gamma$ and
$\Gamma'$ is possible or not, will turn out to be
largely determined by two functions
$f_1^{\Gamma\rightarrow \Gamma'},f_2^{\Gamma\rightarrow \Gamma'}$
that will be called
minimal functions for reasons
that will become clear later.
Let $g:{{\mathbbm{R}}^{+}}^2\times {\mathbbm{R}}^2\rightarrow
{\mathbbm{R}},$
\begin{eqnarray}
    g(a,b,c,d):= 
    (a^2-1)+(b^2 -1 ) c^2 d^2 + 2
    c d- ab(c^{2}+ d^{2}).\nonumber
\end{eqnarray}
For a pair $(\Gamma,\Gamma')$ of
covariance matrices with associated
vectors $(\xi_1, \xi_2 ,\xi_3,\xi_4)$
and $(\xi_1', \xi_2' ,\xi_3',\xi_4')$ with $\xi_{3},\xi_{4}>0$
define the two functions
$f_1^{\Gamma\rightarrow \Gamma'},f_2^{\Gamma\rightarrow \Gamma'} :
{\mathbbm{R}}^+\times {\mathbbm{R}} \rightarrow {\mathbbm{R}}$
as
\begin{eqnarray}
    f_1^{\Gamma\rightarrow \Gamma'}(x,y) &:=&
    g(\xi_{1}',\xi_{1},x/\xi_{3}, y/\xi_{4} ),
    \nonumber\\
    f_2^{\Gamma\rightarrow \Gamma'}(x,y) &:=&
    g(\xi_{2}',\xi_{2},\xi'_{3}/x, \xi'_{4}/y).
    \nonumber
\end{eqnarray}
The first statement concerns LOG in system 1 only.
In this case the criterion amounts to simple
inequalities that have to be satisfied.
The second gives the full result
for general LOG.  

\smallskip

\noindent
{\bf Proposition 1. --} {\it Let $\rho$ and $\rho''$
be Gaussian states of a $1\times 1$-system with
covariance matrices $\Gamma$ and $\Gamma''$
and associated vectors $(\xi_{1},\xi_{2},\xi_{3},\xi_{4})$
and $(\xi''_{1},\xi_{2},\xi''_{3},\xi''_{4})$ with
$\xi_{4},\xi_{4}''>0$.
Then
    $\rho \longrightarrow \rho''\,\,\text{ under LOG}$ in system 1,
if and only if
\begin{eqnarray}\nonumber
    1.\,\,| \xi_{3} \xi_{4}| / \xi_{1}
    \geq
    | \xi''_{3} \xi''_{4}|  /
    \xi''_{1},\,\,\,\,\,\,
    2.\,\,
    f_1^{\Gamma\rightarrow \Gamma''}( \xi_{3}'', \xi''_{4})\geq
    0.
\end{eqnarray}
}
\noindent
{\bf Proposition 2. --} {\it Let $\rho$ and $\rho'$
be Gaussian states of a $1\times 1$-system with
covariance matrices $\Gamma$ and $\Gamma'\,$
and associated vectors $(\xi_{1},\xi_{2},\xi_{3},\xi_{4})$
and $(\xi'_{1},\xi'_{2},\xi'_{3},\xi'_{4})$ with
$\xi_{4},\xi_{4}'>0$.
Then\,\,\,
    $\rho \longrightarrow \rho'\,\,\text{ under LOG}$,
if and only if
one of the points
\begin{eqnarray}\nonumber
    (x,y)\in
    (f_1^{\Gamma\rightarrow \Gamma'})^{-1}(0)
    \cap
    (f_2^{\Gamma\rightarrow \Gamma'})^{-1}(0)
\end{eqnarray}
    satisfies $ |\xi_{3}\xi_{4}| \xi_{1}'/\xi_{1}  \geq |x y| \geq
    |\xi_{3}'\xi_{4}'| \xi_{2}/\xi_{2}'$.
}

\smallskip

\noindent
{\it Proof of Proposition 1.}
We begin with investigating
what conditions have to be met when a
LOG is implemented in system 1 and a symplectic
operation in system 2.
Starting point is a general representation theorem
concerning Gaussian completely positive maps \cite{Dem1}:
Any Gaussian completely positive
map is reflected on the level of
the covariance matrix as a map
\begin{equation}\label{e1}
    \Gamma\longmapsto M^T \Gamma M + G,
\end{equation}
where $M$ and $G$ are real $4\times 4$-matrices,
$G$ is moreover symmetric.
The condition
\begin{eqnarray}\label{e2}
    G+ i \Sigma  - i M^T \Sigma M \geq 0
\end{eqnarray}
on the matrices $M$ and $G$
incorporates the complete positivity of the map.
The state transformation
mapping $\Gamma$ on $\Gamma''$
can be decomposed into three steps:
first, an appropriate
matrix $S=S_1\oplus S_{2} $, $S_1,S_2
\in Sp(2,{\mathbbm{R}})$, is applied on the initial covariance
matrix $\Gamma$, such that
$S^{T}\Gamma S$ is of normal form.
Then a LOG restricted to
system 1 and a symplectic operation in
system 2 is implemented,
mapping $S^{T}\Gamma S$ onto another
matrix in normal form. Finally,
$T=T_1\oplus T_{2}$, $T_1,T_2
\in Sp(2,{\mathbbm{R}})$, is used in order to
transform the resulting matrix into $\Gamma''$.
The second step can be represented in the form of Eq.\ (\ref{e1})
with real matrices $M$ and $G$.
Clearly, the composition of the three steps,
    $\Gamma\longmapsto
     (T^T M^T S^T)  \Gamma  (S M T) + T^T G T$
amounts again to a LOG.
Therefore, we can without loss of generality assume that
both $\Gamma$ and $\Gamma''$ are already in normal
form
with associated vectors
$(\xi_{1},\xi_{2},\xi_{3},\xi_{4})$ and
$(\xi_{1}'',\xi_{2}'',\xi_{3}'',\xi_{4}'')$.
The task is then to find appropriate
real matrices $M$ and $G$ as above  such that
$\Gamma''=M^T \Gamma M + G$,
representing a LOG restricted to system 1 and a
symplectic operation in system 2.
Hence, it is required that $M$ and $G$ are of the form
\begin{eqnarray}\label{e3}
M&=&M_1\oplus M_{2},
G=G_1 \oplus {{0}} ,
\end{eqnarray}
where $G_{1}$ is symmetric and $M_2\in Sp(2,{\mathbbm{R}})$, i.e.,
$M_{2}$ satisfies
$M_{2}^{T} \Sigma M_{2}= \Sigma$.
Due to the normal form of $\Gamma$ and $\Gamma''$
we have that $M_{2}^{T} \text{diag}(\xi_{2},\xi_{2})
M_{2}=\text{diag}(\xi_{2},\xi_{2})$, and
it follows that $M_{2}\in SO(2)$. Let us set
$M_{33}= M_{44}=\text{cos}(\theta/2)$, $M_{34}=-\text{sin}(\theta/2)$,
and $M_{43}= \text{sin}(\theta/2)$ with $\theta\in(-2\pi,2\pi]$.
The requirement that $\Gamma''=M^T \Gamma M + G$
implies then
a certain set of equations that has to be satisfied,
connecting the entries of $M_{1}$ and $M_{2}$. An elementary
calculation yields finally
\begin{eqnarray}
    M_{11}&=& (\xi_{3}''/\xi_{3})\text{cos}(\theta/2),\,\,\label{m1}
    M_{22}= (\xi_{4}''/\xi_{4})\text{cos}(\theta/2),\\
    M_{12}&=& -(\xi_{4}''/\xi_{3})\text{sin}(\theta/2),\,\,
    M_{21}=(\xi_{3}''/\xi_{4})\text{sin}(\theta/2).\label{m2}
\end{eqnarray}
Not all such matrices $M=M_{1}\oplus M_{2}$
and $G=\Gamma'' - M^{T}\Gamma M$ define a
completely positive map, however.
Due to the block diagonal form of $M$ and $G$,
the inequality (\ref{e2})
reflecting the complete positivity
can be written as
\begin{eqnarray}\nonumber
H_1:= G_1 + i \Sigma (1- |M_1|)\geq 0.
\end{eqnarray}
As $H_1$ is a
Hermitian $2\times 2$-matrix, $H_{1}\geq 0$ is
in turn equivalent to
$|H_1|\geq 0, \,\text{tr}[H_1]\geq 0$.
The determinant
$|M_{1}|=(\xi_{3}''\xi_{4}'')/(\xi_{3}\xi_{4})$
is independent of $\theta$.
The determinant and trace of $H_{1}$ can be evaluated to
$\text{tr}[H_{1}]= 2 \xi_{1}'' - \xi_{1} \| M_{1} \|^{2}$ and
$|H_{1}|= (\xi_{1}'')^{2} - \xi_{1}\xi_{1}'' \| M_{1} \|^{2}
+ \xi_{1}^{2}
|M_{1}|^{2}-(1-|M_{1}|)^{2}$,
where $\| M_1 \|$ denotes the Hilbert-Schmidt norm of $M_1$.
Hence, it is always optimal to chose $\theta$ in such a way
that
\begin{eqnarray}\nonumber
    \|M_{1}\|^{2}=\Bigl[\frac{(\xi_{3}'')^{2}}{\xi_{3}^{2}}+
    \frac{(\xi_{4}'')^{2}}{\xi_{4}^{2}}\Bigr]\text{cos}^{2}
  \frac{\theta}{2}+
    \Bigl[\frac{(\xi_{4}'')^{2}}{\xi_{3}^{2}}+
    \frac{(\xi_{3}'')^{2}}{\xi_{4}^{2}}
    \Bigr]\text{sin}^{2}\frac{\theta}{2}
\end{eqnarray}
    is minimal.
But since $\xi_{3}^{2}\geq \xi_{4}^{2}$ and
$(\xi_{3}'')^{2}\geq (\xi_{4}'')^{2}$, it is true
that always
$(\xi_{3}''/\xi_{3})^{2}+
   (\xi_{4}''/\xi_{4})^{2}\leq
   (\xi_{4}''/\xi_{3})^{2}+
    (\xi_{3}''/\xi_{4})^{2}$,
    and therefore, $\theta=0$ is the optimal choice.
    To simplify the structure of the requirements one can proceed
    as follows:
    The inequality
$|H_1|\geq 0$ implies in particular that
\begin{equation}\label{dn0}
    \xi_1''/\xi_1
    \geq |{\xi_{3}'' \xi_{4}''}|/|\xi_3 \xi_4|.
\end{equation}
Whenever (\ref{dn0}) is satisfied, $|H_{1}|\geq 0$ yields
a stronger upper bound for $\|M_{1}\|^{2}$ as $\text{tr}[H_{1}]\geq 0$
does, as then
\begin{eqnarray}\label{later}
    \bigl((\xi_1'')^{2}+ \xi_1^{2} |M_{1}| -
    (1-|M_{1}|)^{2} \bigr) /(\xi_1'' \xi_1)\leq  2 \xi_1''/\xi_1.
\end{eqnarray}
Altogether, this implies that equivalently to
requiring the validity of both $|H_{1}|\geq 0$
and $\text{tr}[H_{1}]\geq 0$
one may require
that both Eq.\ (\ref{dn0}) and  $|H_{1}|\geq 0$
hold.
Therefore, we finally
arrive at the statement that $\rho\longrightarrow \rho''$
under under a LOG in system 1 and a symplectic operation in system 2
if and only if both $
|{\xi_{3}'' \xi_{4}''}|/\xi_1'
    \leq   |\xi_3 \xi_4|/ \xi_1
$ and
    $f_{1}^{\Gamma\rightarrow \Gamma''}(\xi_{3}'',\xi_{4}'')\geq 0$
are satisfied. This criterion depends only on
the invariants with respect to
local symplectic operations, and hence, we
arrive at Proposition 1. \proofend

It is worth noting what physical situation
is reflected by equality $f_{1}^{\Gamma\rightarrow \Gamma''}(\xi_{3}'',\xi_{4}'')= 0$.
Under the constraint that the
entries of $M$ and $G$ are given by
Eqs.\ (\ref{e3},\ref{m1},\ref{m2}),
it can be shown easily that
equality holds if and only if
$M$ and $G$ satisfy
$G= K^T G^{-1} K$,
where $K:= M^T \Sigma M -\Sigma$.
Solutions of this type
are the minimal solutions in the sense of
Ref.\ \cite{Lindblad}:
For a given initial covariance matrix $\Gamma$
and a given matrix $M$
the equation $G= K^T G^{-1} K$
specifies those symmetric
matrices $G$ that add minimal noise.
Such LOG will consequently
be called minimal.
Hence, a LOG in system 1 from $\rho$ to $\rho''$
is minimal
if and only if $f^{\Gamma\rightarrow \Gamma''}_1(\xi_{3}'', \xi_{4}'')= 0$
holds (see Fig.\ 2).
Therefore, one may interpret the conditions of
Proposition 1 in physical terms as follows:
the first condition requires that the `stretching'
$|M|$ is sufficiently small, the second
makes sure that enough noise is introduced in the
course of the
transformation.

{\it Proof of Proposition 2.}\/
A general LOG can again be decomposed into
several steps.  As before, without loss of generality one
may assume that the initial and the final covariance
matrices $\Gamma$ and $\Gamma'$ are of normal form
with associated vectors $(\xi_{1},\xi_{2},\xi_{3},\xi_{4})$
and $(\xi'_{1},\xi'_{2},\xi'_{3},\xi'_{4})$, respectively.
In two intermediate steps one transforms
$\Gamma\mapsto\Gamma''$ and
$\Gamma''\mapsto \Gamma'$ by means of
LOG restricted to
one system and appropriate symplectic operations in
the other system. The vector associated with
$\Gamma''$ will be denoted as
$(\xi'_{1},\xi_{2},x,y)$.
One can proceed as before, and after applying
analogous steps one
finally arrives at the criterion that
$\rho\longrightarrow\rho'$ under LOG
if and only if there exists an
$(x,y)\in{\mathbbm{R}}^{+}\times {\mathbbm{R}}$
such that the inequalities
\begin{eqnarray}\label{nf1}
    w_{1}
    \geq | x y | \geq w_{2},\,\,
    f_{1}^{\Gamma\rightarrow \Gamma'}(x,y)&\geq& 0,\,\,
    f_{2}^{\Gamma\rightarrow \Gamma'}(x,y)\geq 0
\end{eqnarray}
are simultaneously satisfied, where
$w_{1}:=|\xi_{3} \xi_{4}|  \xi_{1}'/ \xi_{1}$ and
$w_{2}:=
| \xi_{3}' \xi_{4}' | \xi_{2}/\xi_{2}'$.
This is already a criterion
of its own, but it still requires a search in a two-dimensional
set. The key observation in a simplification
is that the intersection of the interior of the
set $(f_{i}^{\Gamma\rightarrow
\Gamma'})^{-1}(0)$ and and the set
$N_{i}:=\{(x,y)\in {\mathbbm{R}}^{+}\times
{\mathbbm{R}}| |xy|= w_{i}\}$ is empty
for both $i=1,2$ (see Fig.\ 2).
The minimal value of $(x/\xi_{3})^{2}+ (y/\xi_{4})^{2}$
for $(x,y)\in N_{1}$ is given by
$2 \xi_{1}'/\xi_{1}$. Hence,
it follows
from Eq.\ (\ref{later})
that $f_{1}(x,y) < 0$ for all $(x,y)\in N_{1}$, if
$\xi_{1}\neq \xi_{1}'$, and
 $f_{1}(x,y) \leq  0$ for all $(x,y)\in N_{1}$, if
$\xi_{1} =\xi_{1}'$.
Similarly,
$f_{2}(x,y)\leq 0$ for all $(x,y)\in N_{2}$.
Moreover, $f_{1}^{\Gamma\rightarrow
\Gamma'}$ is continuous on ${\mathbbm{R}}\times
{\mathbbm{R}}^{+} $,
and for $f_{2}^{\Gamma\rightarrow
\Gamma'}$
there exists a continuous continuation on ${\mathbbm{R}}^{+} \times
{\mathbbm{R}}$.
The problem is therefore
reduced to the
subsequent search for intersection points:
there exists an $(x,y)\in
{\mathbbm{R}}^{+} \times {\mathbbm{R}}$ satisfying
(\ref{nf1}) if and only if
there exists a point  $(x,y)\in (f_{1}^{\Gamma\rightarrow
\Gamma'})^{-1}(0)\cap(f_{2}^{\Gamma\rightarrow
\Gamma'})^{-1}(0)$ such that
$w_{1} \geq | x y | \geq w_{2}$.
This is Proposition 2.
In particular, this means
that if the transformation $\rho\longrightarrow \rho'$ is
possible, it can always realized as a composition of
two minimal LOG in system 1 and 2, respectively \cite{Rema}.
\proofend

So far, the
simple case has been omitted
that the initial state $\rho$ has a
covariance matrix $\Gamma$
with associated vector $(\xi_{1},\xi_{2},\xi_{3},\xi_{4})$, where
$\xi_{4}=0$. It turns out that one can proceed as before.
In the notation of Proposition 1 (but with $\xi_{4}=0$),
one arrives at the statement that $\rho\longrightarrow \rho''$
under LOG restricted to system 1, if and only if
both
$\xi_{4}''=0$ and
$(\xi_{3}''/\xi_{3})^{2}\leq ( (\xi_{1}'')^{2}-1)/(\xi_{1} \xi_{1}'')$.
Consequently, in the notation of Proposition 2,
$\rho\longrightarrow\rho'$ under LOG, if and only if
    $\xi_{4}''=0$ and
$(\xi_{3}''/\xi_{3})^{2}\leq  (  (\xi_{2}')^{2}-1) (
(\xi_{1}')^{2}-1) /(\xi_{1}\xi_{1}' \xi_{2} \xi_{2}')$.

\begin{figure}
\centerline{
        \epsfxsize=5.3cm
       \epsfbox{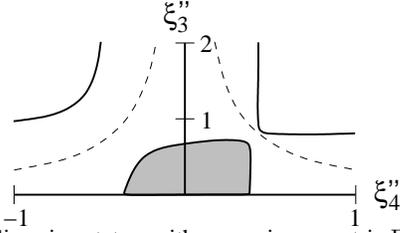}
}
\caption{
Given is a state $\rho$ with a covariance
matrix $\Gamma$
with associated vector $(\xi_1,\xi_2,\xi_3,\xi_4)=
(3,5,1,1/2)$.
The shaded area depicts
what values of
$\xi_{3}''$ and $\xi_{4}''$ are accessible
under a LOG in  system 1,
under the assumption that
the final covariance matrix $\Gamma''$
is associated with a vector
$(\xi_{1}'',\xi_{2},\xi_{3}'',\xi_{4}'')$ with
$\xi_{1}''=2$.
The thick line corresponds to
those points $(\xi_3'',\xi_4'')$ with
$f_1^{\Gamma\rightarrow\Gamma''} (\xi_3'',\xi_4'')=0$
for which the transformation is a minimal LOG,
the dashed line represents the points satisfying
$|\xi_3'' \xi_4''|= |\xi_3 \xi_4| \xi_1''/\xi_1$.}
\end{figure}

As a first application
we can look for Gaussian states $\rho$ and $\rho'$
that are incommensurate, that is, pairs of
states $(\rho,\rho')$
for which neither
$\rho\longrightarrow\rho'$ under LOG nor
$\rho'\longrightarrow \rho$ under LOG holds.
As can readily be verified using Proposition 2,
an example of such a pair
is given by
states specified by covariance
matrices associated with
    $(2,2,1,1)$, $(2,2,1,-1/2)$,
respectively.
The relation that a state can be transformed
into another state under LOG
induces hence
a partial order on the set of
Gaussian states, but
not a total order.

With this letter we have posed and answered
a basic question: Under the
constraint of locality,
we ask which pairs
of Gaussian states allow for a transformation
from one state to the other. The choice for the set
of allowed operations -- Gaussian completely positive maps --
has been motivated by pragmatic
considerations: in quantum optical systems
such operations
can be implemented with present technology.
Needless to say, there are many open questions
that may be approached with similar methods. In particular,
one may take into account selective measurements 
projecting on Gaussian states together with classical communication.
It is the hope that this letter stimulates
such further considerations.

We would like to thank
C.\ Simon, A.\
Winter, G.\ Giedke, 
and K.\ ${\dot {\rm Z}}$yczkowski
for helpful remarks.
This work has been supported by the
European Union
(EQUIP -- IST-1999-11053),
the A.-v.-Humboldt-Stiftung,
and the EPSRC.

\end{multicols}

\begin{references}
\bibitem{MixedRemark}
    The analysis of
    state manipulation of
    mixed states is much 
	more involved and only
    isolated results are known to date \cite{Catalysis}.
\bibitem{LOGd}
	The term Gaussian completely positive map is
	used here as in the mathematical physics literature \cite{Lindblad,Dem1},
	as the generalisation of convolutions of Gaussian measures in 
	the commutative case. This set includes all possible completely
        positive trace-preserving operations that map Gaussian states on
	Gaussian states, under exclusion of those operations that consist
	in a dilation of selective measurements projecting 
	on the improper eigenstates of a canonical coordinate, which are 
	elements of the dual space of the standard test space,
	together with further operations
	depending on the (continuous) measurement outcome.
\bibitem{Ga}
	 A state $\rho$
	is called Gaussian,
	if the characteristic
	function $\phi$, defined as a function of
	${\mathbf{z}}=(x_1,p_1,x_2,p_2)$ as
    $\phi({\mathbf{z}}):=
    \text{tr}[D({\mathbf{z}})\rho]$,
    $D({\mathbf{z}}):=
     \text{exp} (
    i \sum_{n,m=1}^{4} z_n \Sigma_{n m } O_m 
    )$,
is a Gaussian function in phase space.
\bibitem{Inv}
    The four numbers
    $|A_1|$,$|A_2|$, $|B|$, and $|\Gamma|$
    are invariant under all $S\in Sp(4,{\mathbbm{R}})$
    of the form $S=S_1\oplus S_2$.
\bibitem{}
	For completeness we sketch the proof of 
	the equivalence of 
	maps of the form Eqs.\ (\ref{e1},\ref{e2}) and LOG
	as a dilation in the sense 
	of the main text. 
	It is sufficient to show this for each of the
	parts of the composite $1\times 1$-system.
	Let $\Gamma$ and $\gamma$ be covariance matrices of states of a 
	system with one and $N$ canonical degrees of freedom, respectively,
	let $S \in Sp(N+1,{\mathbbm{R}})$, 
	and let $\tilde \Gamma:=
	[S^T (\Gamma\oplus \gamma) S]_{2}$,
	where $[.]_{2}$ 
	denotes the $2\times 2$
	principal submatrix corresponding
	to the original system. Obviously, the resulting
	map is of the form 
	$\Gamma\mapsto \tilde\Gamma = M^T \Gamma M+G$, 
	with $M$, $G$ being
	real $2\times 2$-matrices, and $G$ being symmetric.
	In turn, any such map can be realized 
	with an ancilla system with only $1$ degree
	of freedom: this can be seen when writing $S\in Sp(4,{\mathbbm{R}})$ 
	in block
	form and observing that $A^T\Sigma A + B^T \Sigma B=\Sigma$
	is equivalent with $|A|+|B|=1$ for real $2\times 2$-matrices $A$, $B$.
	Hence, for given real $2\times 2$-matrices $M$, $G$ ($G$ symmetric)
	satisfying  	Eq.\ (\ref{e2}) one can always construct an 
	appropriate $S\in Sp(4,{\mathbbm{R}})$
	and a $2\times 2$ covariance matrix $\gamma$.
\bibitem{Rema}
Practically, it is helpful to consider
explizit parametrizations of
$(f_{1}^{\Gamma\rightarrow \Gamma'})^{-1}(0)$ and
 $(f_{2}^{\Gamma\rightarrow \Gamma'})^{-1}(0)$:
there exist sets $C_{i}$ and two
functions $x_{i}^{\pm}: C_{i}\longrightarrow {\mathbbm{R}^{+}}$
each which satisfy
$(f^{\Gamma\rightarrow \Gamma'}_{i}) ( x_{i}^{\pm}(y),y)=0$ identically.
One has then to look for the intersection
points of these explicit parametrizations.
\end{references}
\end{document}